\documentclass[a4paper,11pt]{article}
\pdfoutput=1 

\usepackage{jcappub} 

\usepackage[T1]{fontenc} 
\usepackage{graphicx}  
\usepackage{dcolumn}   
\usepackage{bm}        
\usepackage{amssymb}   
\usepackage{mathrsfs}  
\usepackage{amsmath}  
\usepackage{gensymb} 
\usepackage[caption=false]{subfig} 
\usepackage{hyperref}

\title{\boldmath Theoretical Estimate of the Sensitivity of the CUORE Detector to Solar Axions}

\author[a,1]{Dawei. Li,\note{Corresponding author.}}
\author[a]{R.J. Creswick,}
\author[a]{F.T. Avignone III,}
\author[b]{and Yuanxu. Wang}

\affiliation[a]{Department of Physics and Astronomy, University of South Carolina, Columbia, SC, USA}
\affiliation[b]{School of Physics and Electronics, Henan University, Kaifeng, Henan, China}

\emailAdd{li255@email.sc.edu}

\abstract{In this paper we calculate the potential sensitivity of the CUORE detector to axions produced in the Sun through the Primakoff process and detected by the inverse coherent Bragg-Primakoff process. The conversion rate is calculated using density functional theory for the electron density and realistic expectations for the energy resolution and background of CUORE.  Monte Carlo calculations for $5~$y$\times741~$kg=$3705~$kg y of exposure are analyzed using time correlation of individual events with the theoretical time-dependent counting rate and lead to an expected limit on the axion-photon coupling $g_{a\gamma\gamma}<3.83\times 10^{-10}~GeV^{-1}$ for axion masses less than $100$ eV.}

\begin{document}
\maketitle
\flushbottom

\section{Introduction}
The CP-violating term in Quantum Chromodynamics(QCD) implies that the neutron electric-dipole moment should be a factor of $10^{10}$ larger than the experimental upper bound \cite{NED}. This CP problem was solved dynamically by Peccei and Quinn \cite{PQ1,PQ2} through the introduction of a global $U(1)_{PQ}$ symmetry. Weinberg \cite{Weinberg} and Wilczek \cite{Wilczek} showed that the axion, which is a Nambu-Goldstone boson, is produced as a consequence of the spontaneously breaking of the $U(1)_{PQ}$ symmetry. Axions, or more generally axion-like particles(ALPs), are pseudo-scalar bosons that can couple with the electromagnetic field or directly with leptons or quarks. The prospect that axions can be a candidate for dark matter in the universe \cite{Preskill,Abbott, Dine, Davis} has motivated many experimental searches \cite{Brookhaven, Japan, SOLAX,DAMA90CL,Japan2, COSME,ADMX,CAST,CASTvacuum,CDMS,EDELWEISS,CASTBufferGas} and theoretical investigations \cite{Creswick, CreswickPRD}. A detailed review of axion physics, astrophysical bounds on the mass and coupling of axions to conventional particles and the cosmological role of axions has been given by Raffelt \cite{Raffelt}.
 
\begin{figure}
    \subfloat[]{{\includegraphics[width=0.5\textwidth]{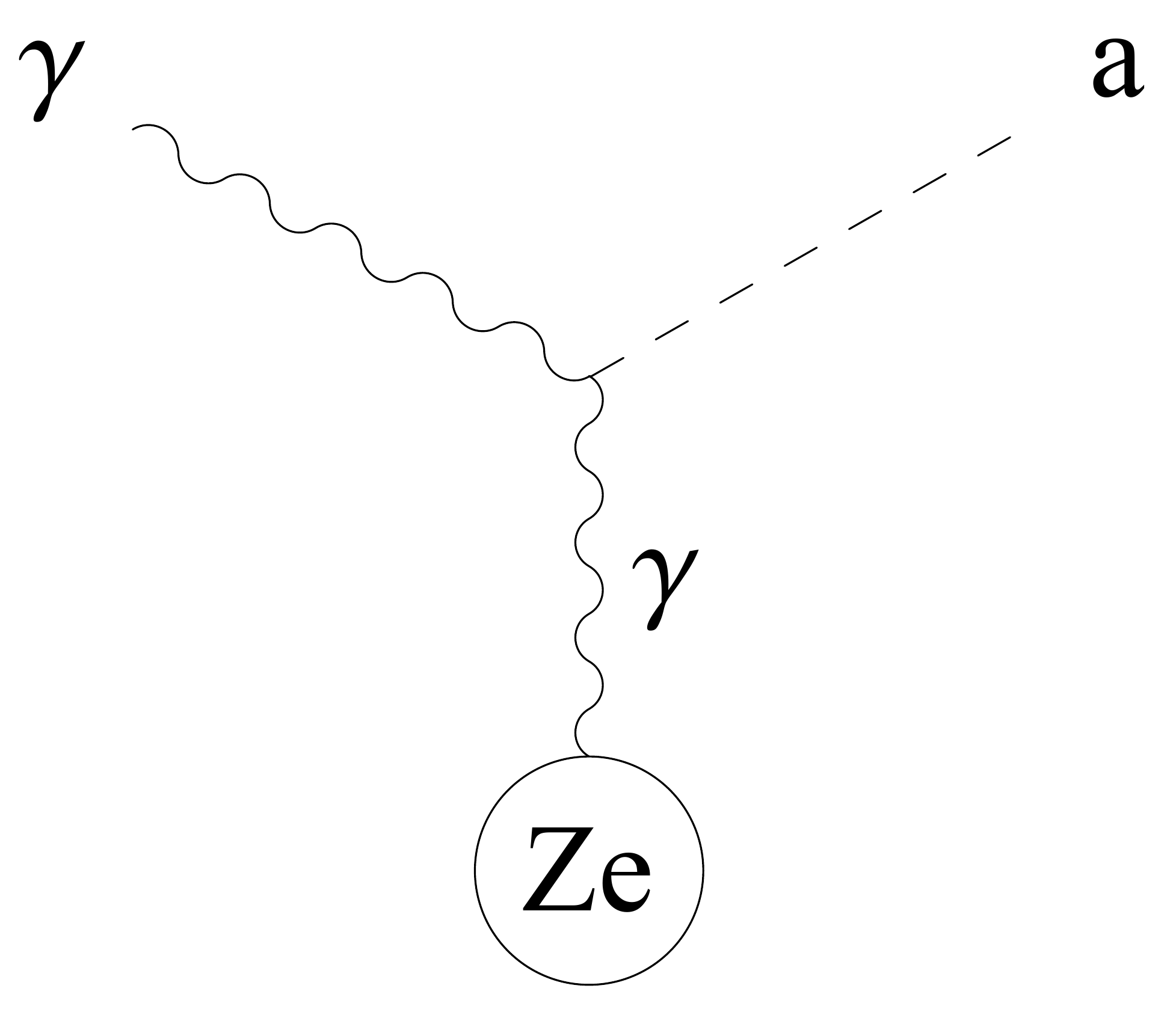} }}
    \subfloat[]{{\includegraphics[width=0.5\textwidth]{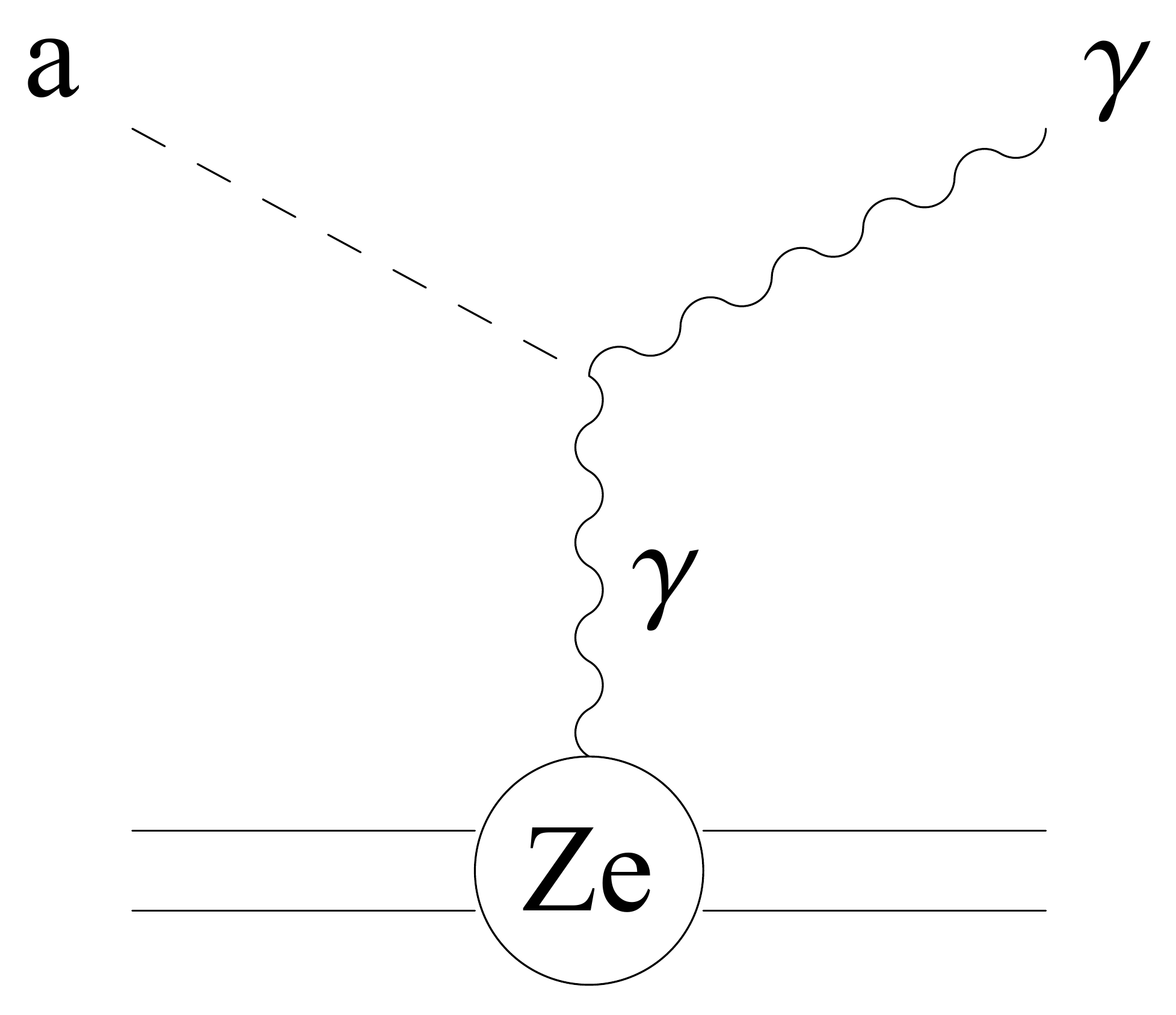} }}
   \caption{\label{fig:Primakoff}(a) An axion is produced in the solar core by the Primakoff effect: A photon couples to a virtual photon in the Coulomb field of the nucleus. (b) An axion couples to a charge in the detector via a virtual photon in the Coulomb field of the crystal producing a photon by the inverse Primakoff effect \cite{SOLAX}.}
\end{figure}

Sikivie \cite{Sikivie} proposed the detection of axions produced in the Sun by the Primakoff process shown in figure~\ref{fig:Primakoff}(a) using a magnetic helioscope. A detailed calculation of the solar axion flux was carried out by van Bibber {\sl et al.} \cite{Bibber}. Buchm\"{u}ller and Hoogeveen \cite{Hoogeveen} proposed using the Primakoff process to produce axions from an intense X-ray source by coherent Bragg conversion in a single crystal and the inverse coherent Bragg-Primakoff process to detect the axions. Paschos and Zioutas \cite{Zioutas} proposed detecting solar axions using inverse Bragg-Primakoff effect, figure~\ref{fig:Primakoff}(b). CUORE(Cryogenic Underground Observatory for Rare Events) \cite{CUORE, CUOREproposal} is a very low background low temperature bolometric detector that is designed to search for neutrinoless double beta decay($0\nu\beta\beta$). It can be sensitive enough to search for dark matter and solar axions. In this paper we calculate the expected sensitivity of CUORE to the upper bound for the coupling of axions to photons via the coherent Primakoff process in $TeO_2$ single crystals.

\section{Theoretical Counting Rates}
The Lagrangian density for the electromagnetic field $\boldsymbol A$ coupled to the axion field $\phi$ is(in natural units $\hbar=c=1$)
\begin{equation}
\mathscr L
=\frac{1}{2}\left(\partial_\mu\phi\partial^\mu\phi-m^2\phi^2\right)
+\frac{1}{2}\left(\left(\partial_t{\boldsymbol A}\right)^2-(\nabla\times\boldsymbol A)^2\right)-\frac{1}{M}\boldsymbol E\cdot\boldsymbol B\phi
\end{equation}
where $\boldsymbol A$ is a vector potential, $1/M\equiv g_{a\gamma\gamma}$ is the axion-photon coupling constant and $\phi$ is the axion field.

The matrix element for a conversion of an axion with momentum $\boldsymbol p$ and energy $E_a$ to a photon with momentum $\boldsymbol k$ energy $E_{\gamma}$ and polarization $\boldsymbol\epsilon$ can be written as:
\begin{equation}
\begin{split}
\mathcal M&=\langle\boldsymbol k\boldsymbol\epsilon;0|\mathcal H_{int}|0;\boldsymbol p\rangle \\
&=\frac{1}{V}\frac{\sqrt\alpha}{Mc^2}\frac{\hbar^3c^3}{2E_a^2}\frac{\boldsymbol\epsilon\cdot(\boldsymbol p\times\boldsymbol k)}{(\boldsymbol p-\boldsymbol k)^2}\tilde\rho(\boldsymbol p-\boldsymbol k)\delta(E_a-E_{\gamma}) 
\end{split}
\end{equation}
where $V$ is the volume considered and $\tilde\rho$ is the Fourier transform of the charge density distribution in units of the fundamental charge $e$. In a periodic lattice $\tilde\rho(\boldsymbol p-\boldsymbol k)$ vanishes unless $\boldsymbol p-\boldsymbol k=\boldsymbol G$, a reciprocal lattice vector, which means that the momentum transfer $\boldsymbol q$ must be equal to $G=2\pi(\frac{h}{a},\frac{k}{b},\frac{l}{c})$, where a, b and c are lattice constants, h, k, l are integers. The $TeO_2$ crystal used by CUORE has a tetragonal symmetry with space group $D_4(422)$ and can be treated as a distorted rutile structure with $a=b=4.8088~$\AA~and $c=7.6038~$\AA~\cite{TeO2}. We calculated the charge density distribution $\rho(\boldsymbol r)$ for $TeO_2$ crystal based density functional theory \cite{DFTtheorem, DFTequation} by utilizing the WIEN2k package \cite{WIEN2k}. $\tilde\rho(\boldsymbol G)$ is calculated by
$$\tilde\rho(\boldsymbol G)=\int_V\rho(\boldsymbol r)e^{-i\boldsymbol G\cdot\boldsymbol r}d^3r$$

Table.~\ref{tab:table1} shows the comparisons between two different calculation methods, density functional theory(DFT) and screening length \cite{Creswick}. It can be seen that $|\tilde\rho(\boldsymbol G)|^2$ given by the screening length is bigger than that given by DFT, which means that the screening length method gives us a higher conversion rate and thus a more strict bound on the coupling constant $g_{a\gamma\gamma}$.

\begin{table}
\begin{center}
\caption{\label{tab:table1}Selected reciprocal lattice vectors that contribute to the inverse Primakoff conversion of solar axions in $TeO_2$.  }
\begin{tabular}{|c|c|c|c|c|c|}
\hline
$(h,k,l)$\footnotemark[1]   &   $d$(\AA)\footnotemark[2]    & $E_0$(keV)\footnotemark[3]   &    $mult$\footnotemark[4] & $|\tilde\rho_c^{DFT}(\boldsymbol G)^2|$\footnotemark[5] & $|\tilde\rho_c^{SL}(\boldsymbol G)^2|$\footnotemark[6]\\
\hline
(1,1,1) & 3.0889 & 2.01 & 8 & 72.53     & 118.98\\
(2,2,1) & 1.6536 & 3.75 & 8 & 505.88   &1021.16\\
(1,2,3) & 1.6265 & 3.81 & 16 & 220.69     &663.13\\
(4,2,0) & 1.0723 & 5.78 & 8  & 10620.90 & 29125.90\\
(3,2,3) & 1.1737 & 5.28 & 16 & 481.43  &1988.35\\
(3,4,1) & 0.9513 & 6.52 & 16 & 1280.11 & 3818.14\\
(3,3,3) & 1.0296 & 6.02 & 8  & 435.17  & 2787.61\\
(2,3,4) & 1.0842 & 5.72 & 16 & 363.21 & 114.08\\
(1,5,2) & 0.9121 & 6.80 & 16 & 24.71  & 80.38\\
(4,2,4) & 0.9304 & 6.66 & 16 & 9969.21 & 30530.90\\
(2,3,5) & 0.9944 & 7.39 & 16 & 549.24  &2118.81\\
(5,4,1) & 0.7452 & 8.32 & 16 & 1641.36 & 5967.68\\
(5,3,3) & 0.7811 & 7.94 & 16 & 2411.45 & 5240.20\\
(1,3,6) & 0.9635 & 6.43 & 16 & 15.98 & 30.00\\
(3,4,5) & 0.8076 & 7.68 & 16 & 3339.10 & 3984.78\\
(1,5,5) & 0.7964 & 7.78 & 16 & 2271.45 & 4415.45\\
(6,1,4) & 0.7265 & 8.53 & 16 & 243.03 & 117.57\\
(3,3,6) & 0.8377 & 7.40 & 8  & 47.86 & 273.31\\
(7,2,2) & 0.6487 & 9.56 & 16 &13981.86 & 23033.10\\
(6,0,5) & 0.7051 & 8.79 & 8  & 984.16 & 6131.84\\
(2,3,7) & 0.8335 & 7.44 & 16 & 1540.88 & 2226.21\\
(1,1,8) & 0.9021 & 6.87 & 8 & 11543.29 & 36814.40\\
(4,4,6) & 0.7012 & 8.84 & 8 & 238.77 & 855.70\\
(7,2,4) & 0.6514 & 9.98 & 16 & 765.01 & 622.35\\ 
(6,1,6) & 0.6665 & 9.30 & 16 & 9789.80 & 27343.1\\
(7,0,5) & 0.6230 & 9.95 & 8 & 1261.85 & 8147.2\\
(6,3,6) & 0.6203 & 9.99 & 16 & 14051.54 & 25538.9\\
\hline
\end{tabular}
\end{center}
\end{table}

\footnotetext[1]{The integers $(h,k,l)$ are the components of reciprocal lattice vectors $\boldsymbol G=2\pi(\frac{h}{a}, \frac{k}{a}, 
\frac{l}{c})$.}
\footnotetext[2]{$d$ is the distance between Bragg planes for a given $\boldsymbol G$ and $d=2\pi/G$.}
\footnotetext[3]{$E_0$ is the minimum energy for which a zero rest mass particle can Bragg scatter with momentum transfer $\boldsymbol G$.}
\footnotetext[4]{$mult$ is the multiplicity, or the number of reciprocal lattice vectors in each family of planes.}
\footnotetext[5]{$|\tilde\rho_c^{DFT}(\boldsymbol G)^2|$ is the Fourier transform of the charge density distribution $\rho$ calculated by WIEN2k based on density functional theory.}
\footnotetext[6]{$|\tilde\rho_c^{SL}(\boldsymbol G)^2|$ is Fourier transform of the charge density distribution $\rho$ based on screening length \cite{Creswick}.}

\begin{figure}
\begin{center}
\includegraphics[scale=0.45]{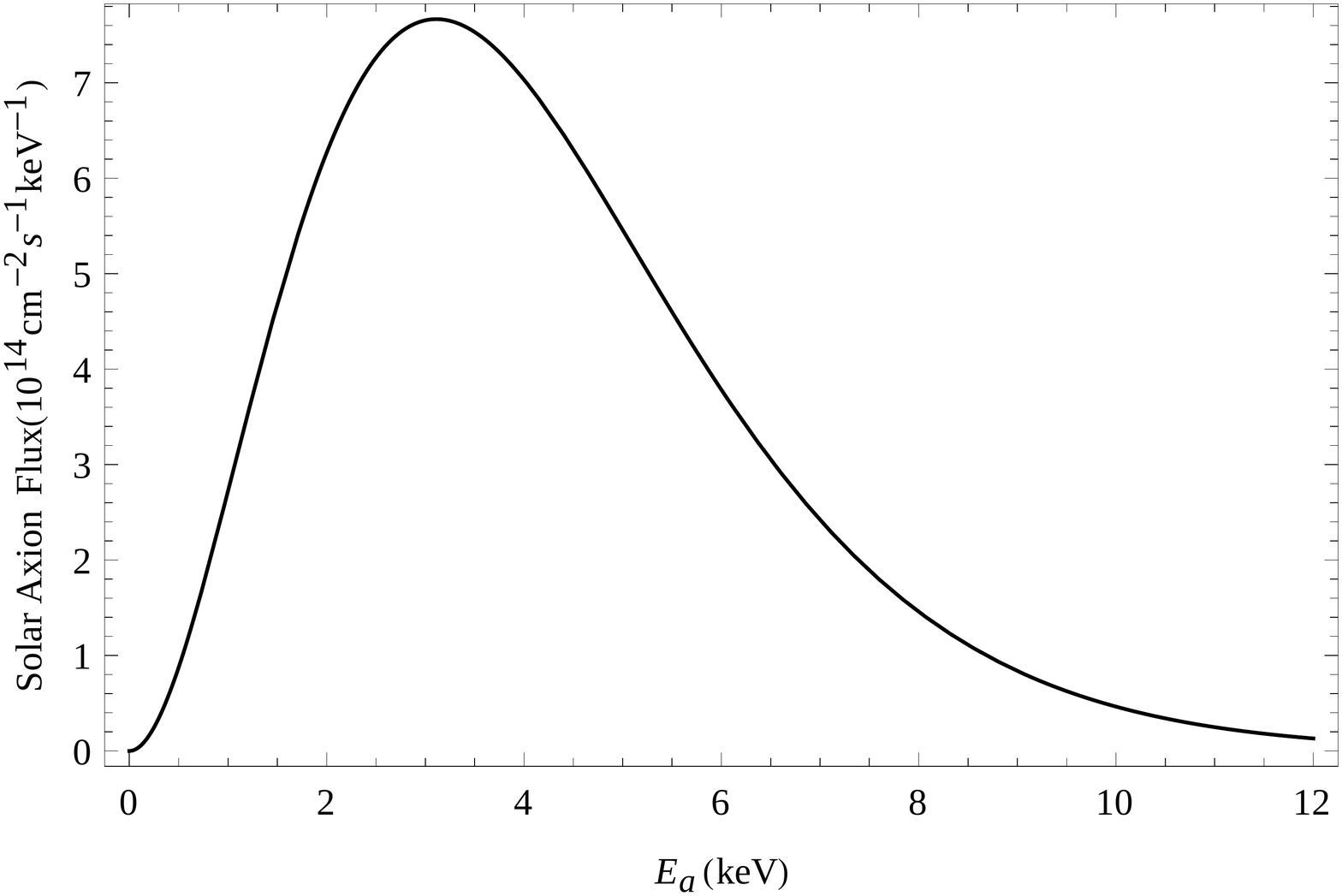}
\caption{\label{fig:axionflux}Solar Axion Flux with $\lambda=1$($g_{a\gamma\gamma}=10^{-8} GeV^{-1})$}
\end{center}
\end{figure}

Integrating over all photon final states, the lowest-order of the conversion rate becomes
\begin{equation}
T(\boldsymbol p)=\frac{4c\pi^2\alpha N_c}{v_c V}\left(\frac{\hbar}{Mc}\right)^2\sum_{\boldsymbol G}|\tilde\rho_c(\boldsymbol G)|^2\frac{|\boldsymbol p\times\boldsymbol G|^2}{G^6}\delta(E_a-E_{\gamma})
\end{equation}
where $v_c$ is the volume of the conventional unit cell, $N_c$ is the number of unit cells. The cross section is related to the conversion rate by $T(\boldsymbol p)=\Phi(\boldsymbol p)\sigma(\boldsymbol p)$ and the flux of a single axion is $\frac{v_a}{V}$, where $v_a$ is the speed of the axion. For very light axions, $v_a\approx c$, yielding
\begin{equation}
\sigma(\boldsymbol p)=\frac{4\pi^2\alpha N_c}{v_c}\left(\frac{\hbar}{Mc}\right)^2\sum_{\boldsymbol G}\left|\frac{\tilde\rho_c(\boldsymbol G)}{G^2}\right|^2\frac{|\boldsymbol p\times\boldsymbol G|^2}{G^2}\delta(E_a-E_{\gamma})
\label{Eq:sigma}
\end{equation}
The flux of axions from the Sun has been calculated by van Bibber {\sl et al.} \cite{Bibber} and can be well approximated by the empirical form
\begin{equation}
\frac{d\Phi}{dE}=\sqrt\lambda\frac{\Phi_0}{E_0}\varphi(E/E_0)
\end{equation}
where 
\begin{equation}
\lambda=(\frac{10^8GeV}{Mc^2})^4=(g_{a\gamma\gamma}\times 10^8 GeV)^4
\label{eq:lambda}
\end{equation}
is a dimensionless parameter which uses $M=10^8GeV$ as a benchmark, $\Phi_0=5.95\times 10^{14}~cm^{-2}s^{-1}$ and $\varphi(E/E_0)=\frac{(E/E_0)^3}{\exp(E/E_0)-1}$. When helium and metal diffusion are included, the core temperature of the solar model will be changed a little bit. To take into account this small change, we use the adjusted value of $E_0=1.103~keV$ \cite{Creswick}. Figure~\ref{fig:axionflux} shows the solar axion flux due to the Primakoff process with $\lambda=1$. 
For convenience, a good order of magnitude for the counting rate can be obtained from the combination of factors
\begin{equation}
\frac{d\dot N_0}{dE}=N_A\frac{\Phi_0}{E_0}(\frac{\hbar}{Mc})^2\lambda=1.12\lambda /keV/d
\end{equation}
where $N_A$ is the Avogadro's number. The number of unit cells can be expressed in terms of the mass of the detector m, and molar mass of the unit cell, $\mu_c$
$$N_c=\frac{m}{\mu_c}N_A$$
Coherent conversion of axions to photons is possible when the energy of the axion and direction to the Sun, $\boldsymbol{\hat p}$, satisfies the Bragg condition,
\begin{equation*}
E(\boldsymbol{\hat p}, \boldsymbol G)=\hbar c\frac{G^2}{2\boldsymbol{\hat p}\cdot\boldsymbol G}
\end{equation*}
Taking into account the fact that the detector has a certain energy resolution, we replace the delta function in eq.~\eqref{Eq:sigma} with a Gaussian function $W_{\Delta}$ with the same full width at half maximum(FWHM) as the detector, and finally we have the conversion rate
\begin{equation}
\begin{split}
\frac{d\dot N}{dE}=&m\hbar c\frac{d\dot N_0}{dE}\frac{4\pi^2\alpha}{\mu_cv_c}\sum_{\boldsymbol G}|\tilde\rho_c(\boldsymbol G)|^2\frac{|\boldsymbol p\times\boldsymbol G|^2}{G^6} \\
& \varphi[E(\boldsymbol{\hat p},\boldsymbol G)/E_0]W_{\Delta}[E-E({\boldsymbol{\hat p},\boldsymbol G})]
\end{split}
\end{equation}
CUORE will have a characteristic low-energy resolution with FWHM=$0.73$ keV at $4.7$ keV and a low background counting rate \cite{BGrate}. Finally, we integrate the total counting rate over a range of energies of width $\Delta E$=$0.5$ keV,
\begin{equation}
R(\boldsymbol{\hat p}, E)=\int_{E'}^{E'+\Delta E}\frac{d\dot N}{dE}(\boldsymbol{\hat p}, E')dE'
\label{eq:axioncountingrate}
\end{equation}

\begin{figure*}
    \centering
    \subfloat{{\includegraphics[width=0.5\textwidth]{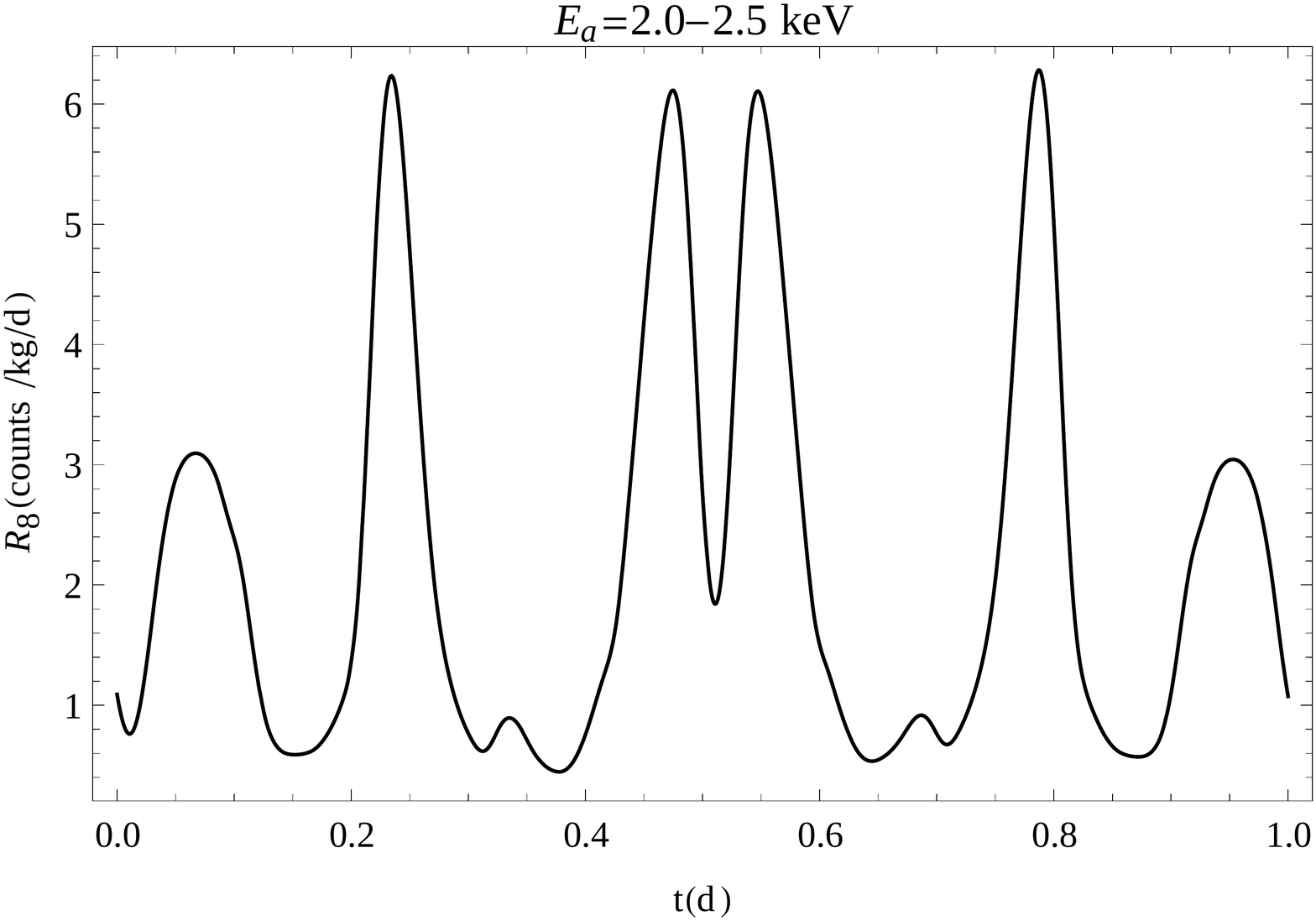} }}
    \subfloat{{\includegraphics[width=0.5\textwidth]{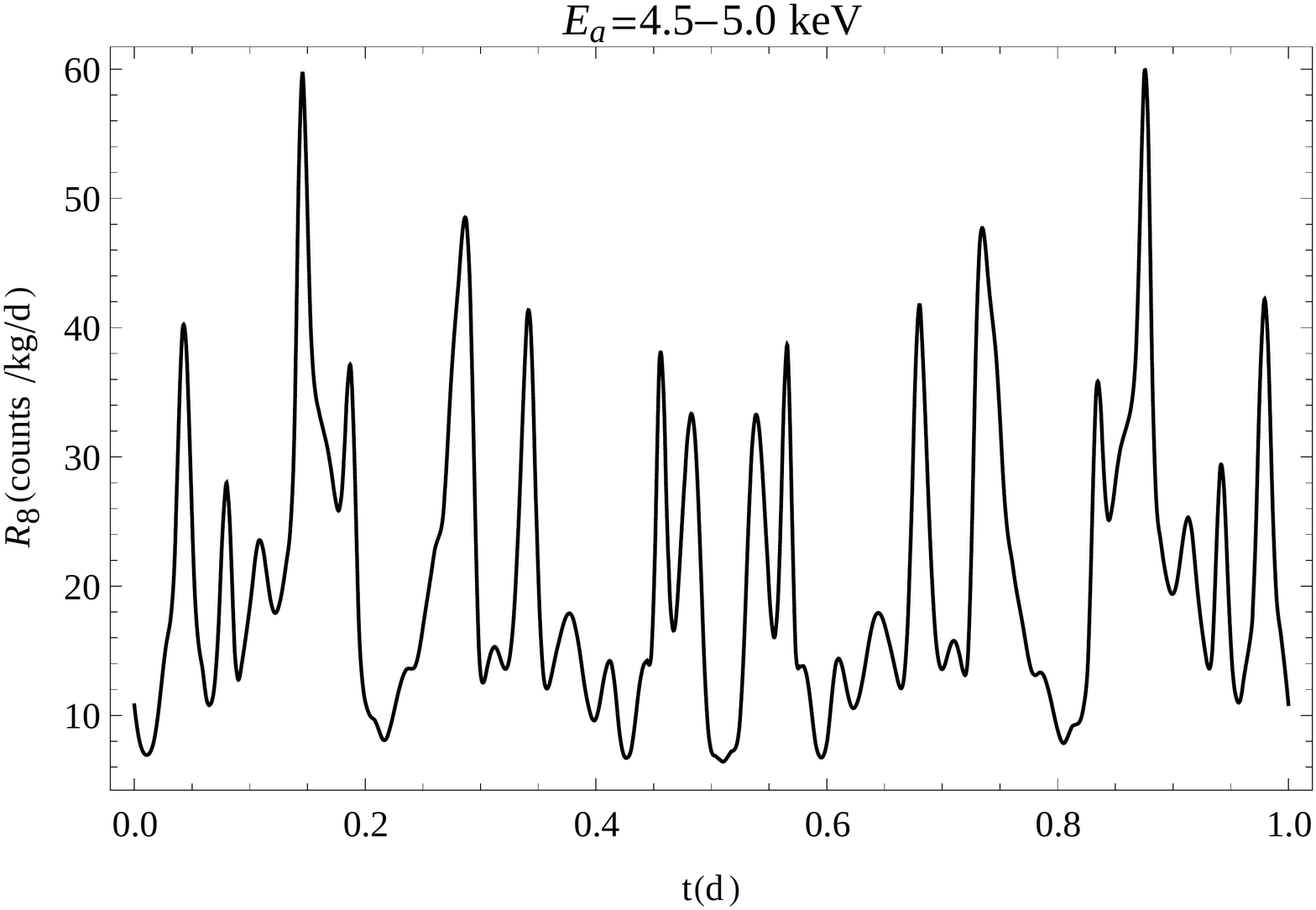} }}
 \qquad
    \subfloat{{\includegraphics[width=0.5\textwidth]{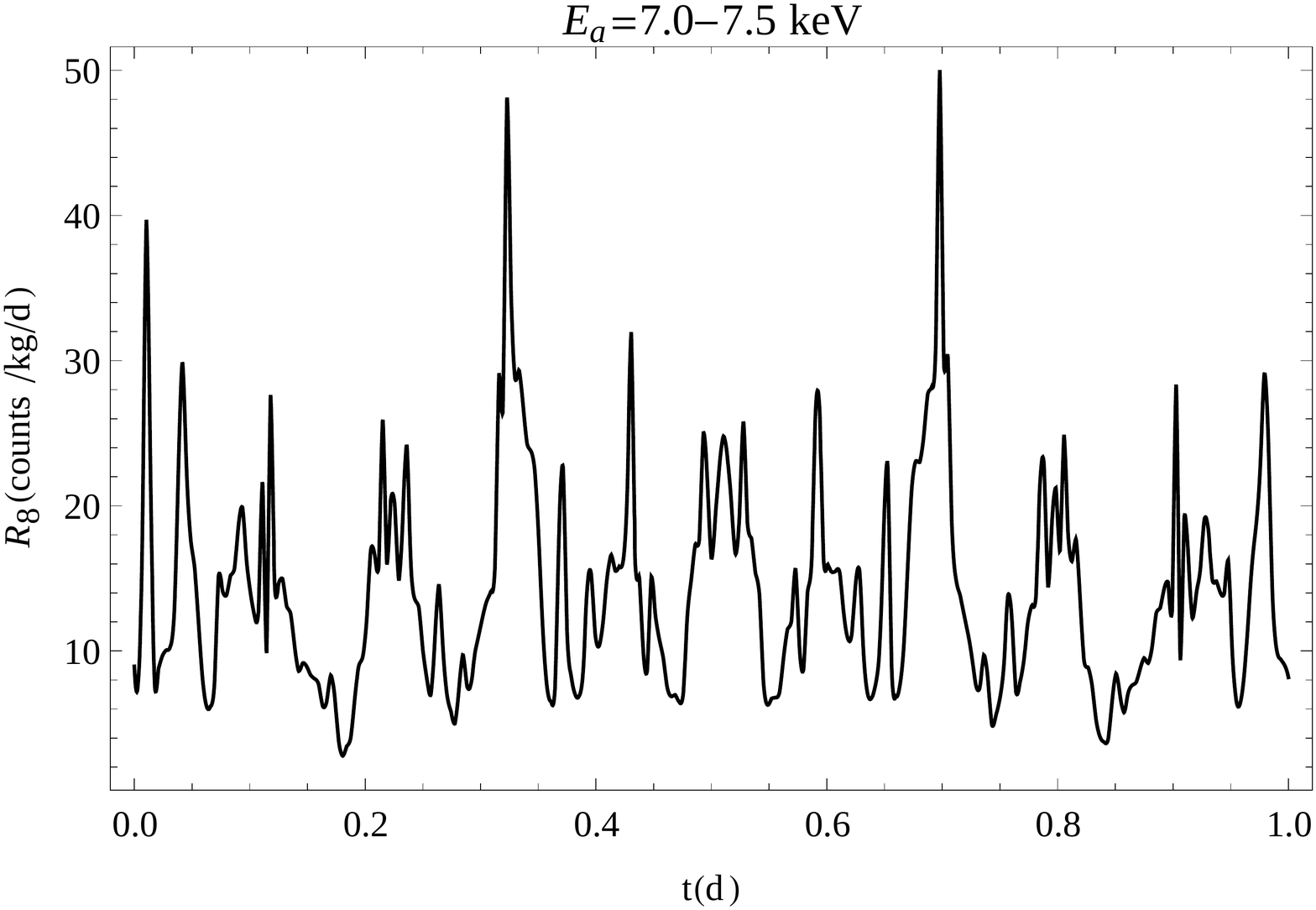} }}
    \subfloat{{\includegraphics[width=0.5\textwidth]{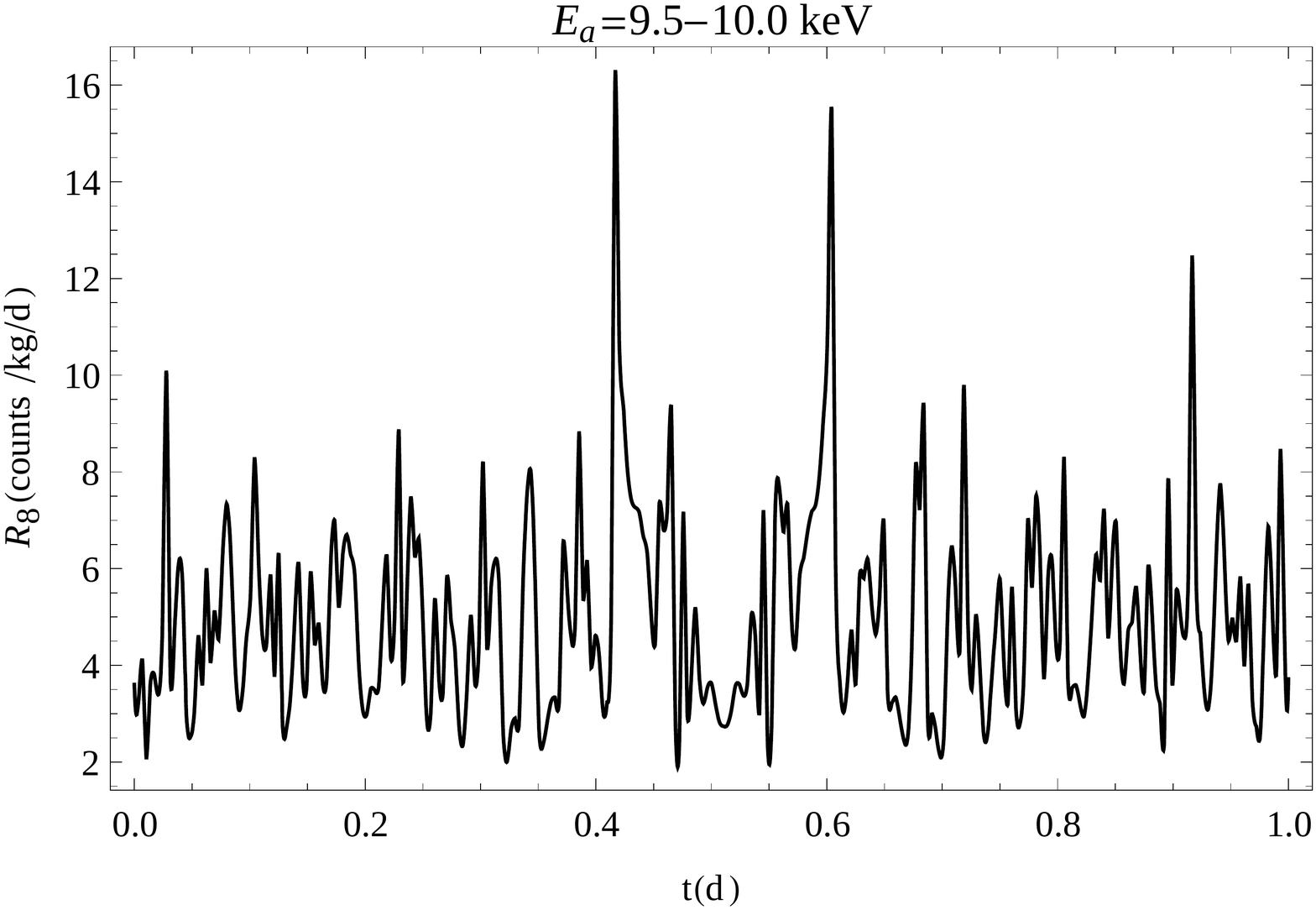} }}
    \caption{\label{fig:countingrate}Expected counting rates $R_8(E, t)$ of photons produced by the inverse Primakoff conversion of solar axions satisfied by the Bragg condition in the CUORE detector, which is located at the Laboratori Nazionali del Gran Sasso(LNGS) in central Italy($42\degree 28'$N $13\degree 33'$E). The rates were calculated for $g_{a\gamma\gamma}=1/M=10^{-8}~GeV^{-1}$ .}
\end{figure*}

\begin{figure}
\begin{center}
\includegraphics[scale=0.45]{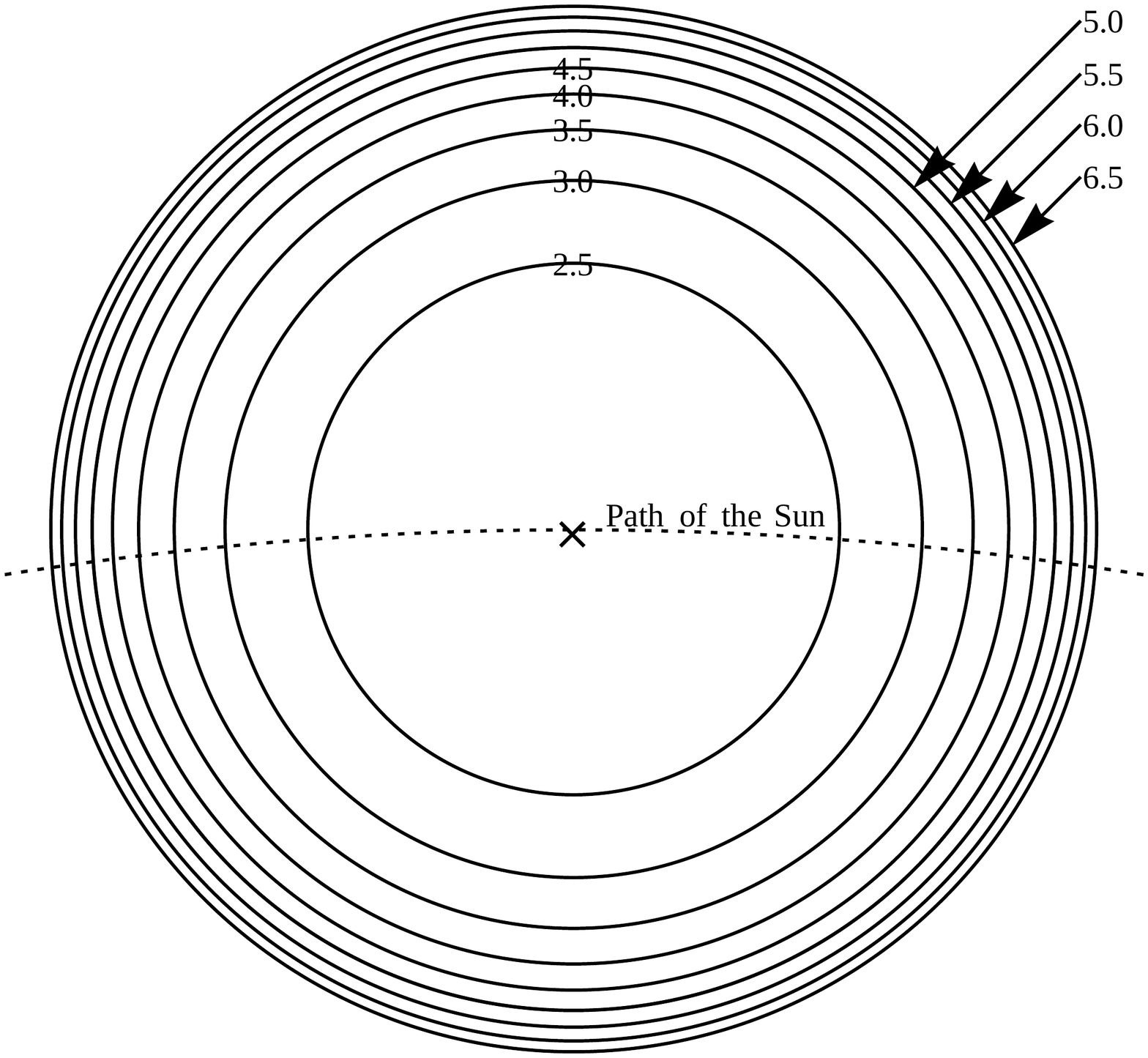}
\caption{\label{fig:bullseyes} Isodetection contours projected on the celestial sphere for the $\boldsymbol G=2\pi(\frac{1}{a},\frac{1}{a},\frac{1}{c})$ plane. The cross sign at the center is the projection of the normal to the (1,1,1) plane. The dotted trojectory represents the path of the Sun through that region.}
\end{center}
\end{figure}

Figure~\ref{fig:countingrate} shows the calculated counting rate as a function of time over a single day for several energy intervals. One way to understand the time-dependent counting rate is that at any instant there might be one or more reciprocal lattice vectors satisfying the Bragg condition. If one considers the contribution to the counting rate of a single $\boldsymbol G$, one can imagine isodetection contours projected on the celestial sphere. Figure~\ref{fig:bullseyes} shows the isodetection contours for axions with energies from $2.5$ keV to $6.5$ keV for $\boldsymbol G=2\pi(\frac{1}{a},\frac{1}{a},\frac{1}{c})$ in steps of $0.5$ keV. The energy bin width is chosen to be slightly bigger than the resolution of the detector. The cross sign at the center is the projection of $\boldsymbol G$ and the dotted trojectory represents a typical path of the Sun through that region. To give the reader some quantitative feeling for the angular size of the isodetection rings, the outermost ring at $6.5$ keV has an angular radius of $72\degree$ and the ring for $6.0$ keV has a radius of $70.5\degree$, so the outermost annulus is $1.5\degree$ wide. The counting rate in the energy bin $6.0-6.5$ keV will rise when the Sun passes through this annulus, which takes about six minutes because the Sun moves $0.25\degree$/min. Then the next annulus with energy $5.5-6.0$ keV will go up and the counting rate in the previous annulus will drop.

From Figure~\ref{fig:countingrate} one can see that the sharpness and complexity of the counting rate increases with energy. This is a geometric effect because, as is clear from Figure~\ref{fig:bullseyes}, the Sun spends less time in the annuli with higher energy than those with lower energy and because there are many more reciprocal vectors available that satisfy the Bragg condition. The daily temporal pattern is dependent on how the Sun pass through the annuli. Figure~\ref{fig:bullseyes} shows the Sun passing along a diameter and through all eight rings, so one could expect the peaks to be seen in all energy bins in a symmetrical pattern. On the other hand the Sun might pass at a grazing trojectory that crosses the outer rings without going through the inner ones, so the patterns shown in one energy bin may not be seen in other energy bins.

\section{Time Correlation Method}

The distinct variation of the counting rates as a function of time suggests that the detection of solar axions can be analyzed using the time correlation method\footnote{Here we suppress the index referencing to a particular energy bin for simplicity}:
\begin{equation}
\chi=\sum_{i=1}^N W(t_i)\times n(t_i) 
\end{equation}
where $n(t_i)$ is a Poisson-distributed random variable at time $t_i$, $W(t_i)$ is the weighting function at time $t_i$, which is subject to the following two constraints

\begin{subequations}
\begin{align}
        \int_0^TW(t)dt&=0\\
        \int_0^TW^2(t)dt&<\infty,
\end{align}
\label{eq:constraints}
\end{subequations}

The expected number of counts in a time interval $\Delta t_i$ is:
\begin{equation}
\langle n(t_i)\rangle=(R_{BG}+\lambda R_8(t_i))\Delta {t_i}
\end{equation}
where $R_{BG}$ is the background counting rate, $R_8(t)$ is the theoretically expected counting rate of solar axions at $g_{a\gamma\gamma}=10^{-8}~GeV^{-1}$ and $\lambda$ is defined in eq.~\eqref{eq:lambda}. Note that $n(t_i)$ is essentially equal to $1$ or $0$ if $\Delta t_i$ is very small. Then the average of $\chi$ is simplified to 
\begin{equation}
\langle\chi(\lambda)\rangle=\lambda\int_0^TW(t)R_8(t)dt
\label{eq:averagechi}
\end{equation}
The variance of $\chi$ becomes
\begin{equation}
\begin{split}
(\Delta\chi(\lambda))^2&=\langle\chi^2\rangle-\langle\chi\rangle^2 \\
&=R_{BG}\int_0^T W^2(t)dt+\lambda\int_0^TW^2(t)R_8(t)dt
\label{eq:chi2}
\end{split}
\end{equation}
Note that the uncorrelated events are canceled out and each event follows Poisson statistics:
\begin{equation}
\langle n^2(t_i)\rangle-\langle n(t_i)\rangle^2=\langle n(t_i)\rangle
\end{equation}
The second term in eq.~\eqref{eq:chi2} is negligible compared with the first term when $\lambda$ is small, so
\begin{equation}
\Delta\chi^2=R_{BG}\int_0^TW^2(t)dt
\end{equation}
The number of events in each time interval is statistically independent, so the probability distribution of $\chi$ given a weighting function $W(t)$ is
\begin{equation}
P(\chi|W)=\left\langle\delta\left(\chi-\sum_{i=1}^NW(t_i)n(t_i)\right)\right\rangle
\end{equation}
The delta function can be represented by its Fourier transform,
\begin{equation}
P(\chi|W)=\int_{-\infty}^{\infty}\frac{1}{2\pi}e^{-i\omega\chi}\prod_{i=1}^N\left\langle e^{i\omega W(t_i)n(t_i)}\right\rangle d\omega
\end{equation}
Expanding the average about $\omega=0$ and keeping the first two terms gives, by eq.~\eqref{eq:averagechi} and eq.~\eqref{eq:chi2}
\begin{equation}
\begin{split}
P(\chi|W)&=\int_{-\infty}^{\infty}\frac{1}{2\pi}\exp\left[-i\omega(\chi-\langle\chi\rangle)-\frac{1}{2}\omega^2\Delta\chi^2\right]d\omega \\
&=\frac{1}{\sqrt {2\pi\Delta\chi^2}}\exp\left[-\frac{(\chi-\langle\chi\rangle)^2}{2\Delta\chi^2}\right]
\end{split}
\end{equation}
which shows that the probability distribution for $\chi$ given the weighting function $W(t)$ is a Gaussian. This is an example of the Central Limit Theorem, which states that data which are affected by many small and unrelated random effects are approximately normally distributed. We want to choose $W(t)$ to maximize $\langle\chi\rangle$ subject to the constraints in eq.~\eqref{eq:constraints}. Using the method of Lagrange multipliers we want to maximize
\begin{equation}
\mathscr F=\langle\chi\rangle-\mu_1\int_0^TW(t)dt-\mu_2\lambda\int_0^TW^2(t)dt
\end{equation}
with respect to $W(t)$, which gives
\begin{equation}
R_8(t)-\mu_1-2\mu_2\lambda W(t)=0
\end{equation}
where $\mu_1$ and $\mu_2$ are multipliers. So
\begin{equation}
W(t)=\frac{1}{2\lambda\mu_2}(R_8(t)-\mu_1)
\label{eq:WFwithfraction}
\end{equation}
$\mu_1$ can be determined by using the constraint that $\int_0^T W(t)dt=0$
$$\int_0^T(R_8(t)-\mu_1)dt=0\Rightarrow\mu_1=\bar R_8$$
where $\bar R_8$ is the average of $R_8(t)$ over the time considered. The second Lagrange multiplier determines the norm of the weighting function. It is convenient to choose $\mu_2=\frac{1}{2\lambda}$ so that the ``best'' weighting function is
\begin{equation}
W(t)=R_8(t)-\bar R_8(t)
\label{eq:weightingfunction}
\end{equation}
The $\log$ likelihood function for $\lambda$ is
\begin{equation}
L(\lambda)\propto -\frac{\left(\chi-\lambda\int W(t)(R_8-\bar R_8)dt\right)^2}{2R_{BG}\int W^2(t)dt}
\label{eq:likelihoodfunction}
\end{equation}
The most probable value for $\lambda$ is
\begin{equation}
\bar\lambda=\frac{\langle\chi\rangle}{\int W^2(t)dt}
\label{eq:barlambda}
\end{equation}
The width of the likelihood function is
\begin{equation}
\Delta\lambda^2=\frac{R_{BG}\int W^2(t)dt}{\left(\int W(R_8-\bar R_8)dt\right)^2}
\label{eq:deltalambda2}
\end{equation}
Taking the variational derivative with respect to $W$ then gives
\begin{equation}
\frac{\partial\Delta\lambda^2}{\partial W(t)}\propto
\frac{W\int W(R_8-\bar R_8)dt-\int W^2(t)dt(R_8-\bar R_8)}{\left(\int W(R_8-\bar R_8)dt\right)^3}
\end{equation}
which vanishes if $W=R_8-\bar R_8$. With this choice of weighting function
\begin{equation}
\Delta\lambda=\sqrt{\frac{R_{BG}}{\int W^2(t)dt}}
\label{eq:deltalambda}
\end{equation}

If we choose the weighting function to be eq.~\eqref{eq:weightingfunction}, not only is $\langle\chi\rangle$ maximized but also $\Delta\lambda$ is minimized. The generalization to the case with several independent energy bins is straightforward; eq.~\eqref{eq:barlambda} becomes
\begin{equation}
\bar\lambda=\frac{\sum_k\langle\chi_k\rangle}{\sum_k\int W_k^2(t)dt}
\label{eq:binbarlambda}
\end{equation}
and eq.~\eqref{eq:deltalambda} becomes
\begin{equation}
\Delta\lambda=\sqrt{\frac{R_{BG}}{\sum_k\int W_k^2(t)dt}}
\label{eq:bindeltalambda}
\end{equation}
where $k$ is the index for the energy bins.

\section{Monte Carlo Simulation}

In a real experiment the total counting rate is given by
\begin{equation}
R=R_{BG}+\lambda R_8
\label{eq:totalrate}
\end{equation}
The background counting rate will dominate when $\lambda$ is small. We can evaluate the sensitivity of the time correlation method by generating pseudo-random data with a given value of $\lambda$ and then evaluating $\bar\lambda$ and $\Delta\lambda$ from eq.~\eqref{eq:likelihoodfunction}, eq.~\eqref{eq:binbarlambda} and eq.~\eqref{eq:bindeltalambda}.

\subsection{Generation of Pseudo-data}
Let $P_0(t,t_0)$ be the probability that no event occurs from $t_0$ to $t$. For a time-dependent counting rate $R(t)$,
\begin{equation*}
P_0(t+\Delta t, t_0)=P_0(t, t_0)\times \left(1-R(t)\times\Delta t\right)
\end{equation*}
which leads to
\begin{equation}
P_0(t,t_0)=e^{-\int_{t_0}^{t}R(t')dt'}
\end{equation}
The probability that the first event takes place at time $t_1$ in a small time interval $\Delta t$ is
\begin{equation*}
P_1(t_1)=e^{-\int_{t_0}^{t_1}R(t')dt'}R(t_1)\Delta t
\end{equation*}

In order to generate a sequence of events with the proper probability distribution, a random number $\mathscr R$ uniformly distributed on the interval $0\leq\mathscr R\leq 1$ is used to determine the time $t$ by solving
\begin{equation}
F(t)=1-e^{-\int_0^t R(t')dt'}
\end{equation}
Note that $F(t)$ also lies in the interval $0\leq F(t)\leq 1$. The probability of choosing a random number $\mathscr R$ in a small interval $\Delta\mathscr R$ is equal to the probability of finding an event at $t_1$ in a small time interval $\Delta t$
$$\Delta\mathscr R=\frac{dF}{dt}\Delta t$$
Differentiating $F(t)$ gives 
$$\Delta\mathscr R=P_0(t_1, 0)R(t_1)\Delta t=P_1(t_1)\Delta t$$
Hence the time of the first event will be distributed correctly by solving $F(t_1)=\mathscr R$. At this point the time is reset and the procedure repeated until we reach the end of the simulation. The times $\{t_i, i=1, \cdots, N\}$ are used to calculate $\chi=\sum_{i=1}^NW(t_i)$, and from this we extract $\bar\lambda$ and $\Delta\lambda$ as described above.

\section{Conclusions}
We have carried out Monte Carlo calculation using the mass, energy resolution and realistic background for the CUORE detector operating for $5$ years. The results show that the CUORE detector with $741$ kg $TeO_2$ in operation for $5$ years can set an upper bound on $\lambda$ of
\begin{equation}
\lambda< 2.15\times10^{-6}
\end{equation}
which is equivalent to an upper limit on the axion-photon coupling constant $g_{a\gamma\gamma}< 3.83\times 10^{-10}$ at $95\%$ confidence level. To illustrate the resolving power of the time correlation method, in five years with $g_{a\gamma\gamma}= 3.83\times 10^{-10}$ there are approximately $600$ events due to axion conversion and $5.5\times10^5$ background events.

\begin{figure}
\begin{center}
\includegraphics[scale=0.5]{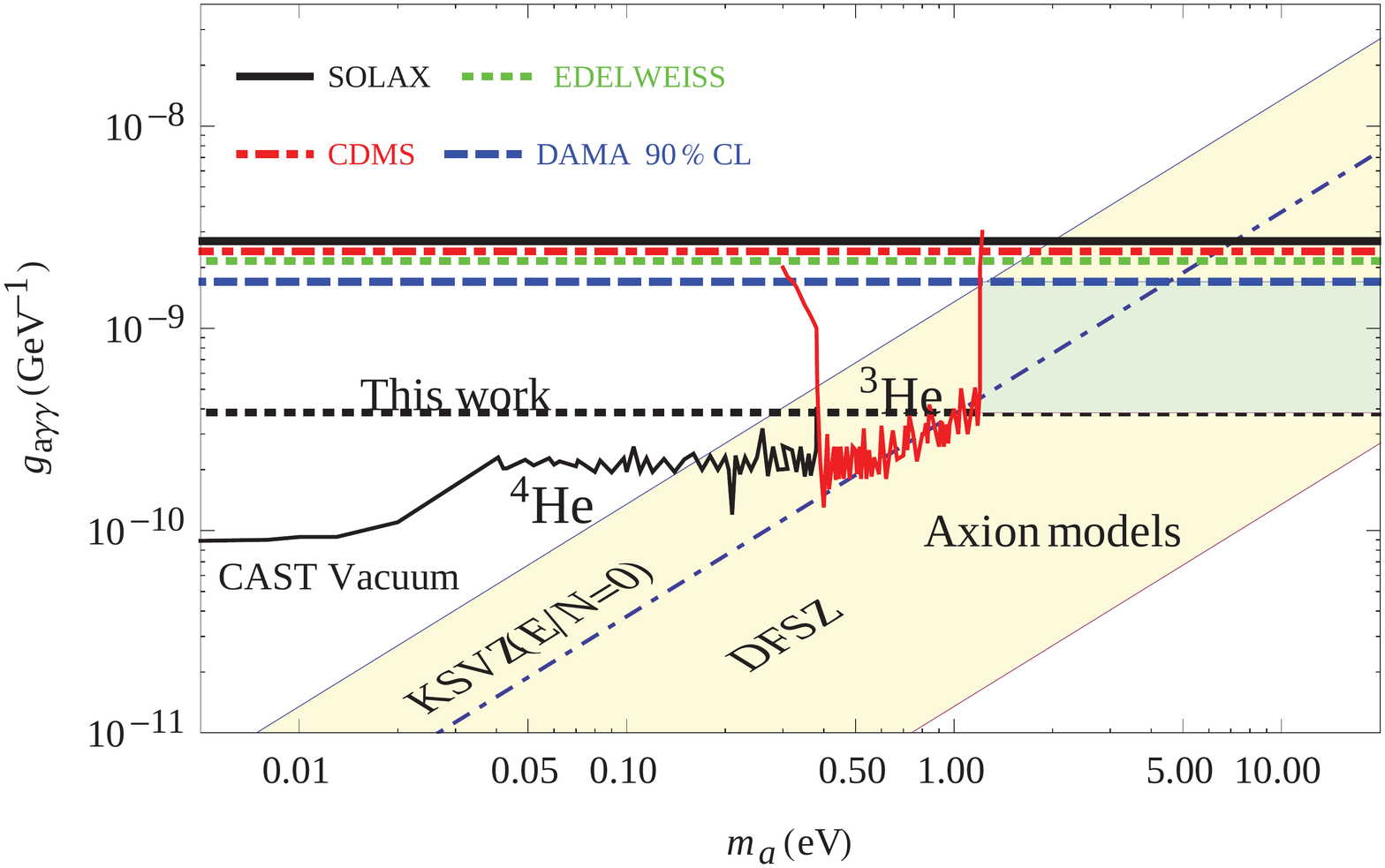}
\caption{\label{fig:axionmodel} Exclusion limits on the $g_{a\gamma\gamma}$-$m_a$ plane. The shaded area is favored by the KSVZ \cite{KSVZ,KSVZ1} and the DFSZ \cite{DFSZ,DFSZ1} axion models. The dotted line shows that with $3705~$kg y of data, CUORE could exclude axions with $g_{a\gamma\gamma}>3.83\times 10^{-10}~GeV^{-1}$ and masses less than $100$ eV.}
\end{center}
\end{figure}

Figure~\ref{fig:axionmodel} is an exclusion plot comparing this calculation with the best limits set by CAST \cite{CAST,CASTvacuum,CASTBufferGas} on the $g_{a\gamma\gamma}$-$m_a$ plane. The lightly shaded area and dotdashed line correspond to various theoretical axion models \cite{KSVZ,KSVZ1, DFSZ, DFSZ1}. Our predicted bound is comparable to the newest CAST results for axions with mass less than $1.2$ eV \cite{CASTBufferGas} and will improve the bound for axion masses in the range $1~ eV \leq m_a\leq~100~eV$ \footnote{The upper limit of $100$ eV is somewhat arbitrary and conservative. The Bragg conversion probability is not very sensitive to axion masses less than $100$ eV, and solar axion flux also varies very little. For axion masses of several hundred eV the solar axion spectrum is distorted and decoherence begins to affect the conversion probability.}, indicated by the darker shaded region(green in color).

Recently, the International Axion Observatory (IAXO), a new generation axion helioscope searching for solar axions by Primakoff conversion in a strong magnetic field, has been proposed (see recent work by J. K. Vogel {\sl et al.} \cite{IAXO}). The predicted sensitivity of IAXO to the coupling constant $g_{a\gamma\gamma}$ is predicted to be on the order of $4\times10^{-12}~GeV^{-1}$ for axion masses less than $0.1$ eV. This is a great improvement over all current experiments, narrowing down search region for axions and dark matter significantly. However, this excluded region of parameter space will not reach beyond $0.2$ eV.

\acknowledgments

This work was supported by the US National Science Foundation Grant PHY-1307204.


\begin{thebibliography}{99}

  \bibitem{NED}
C. A. Baker {\sl et al.}, Phys. Rev. Lett. {\bf 97}, 131801  (2006).

 \bibitem{PQ1}
R. D. Peccei and H. R. Quinn, Phys. Rev. Lett. {\bf 38}, 1440  (1977).

 \bibitem{PQ2}
R. D. Peccei and H. R. Quinn, Phys. Rev. D {\bf 16}, 1791  (1977).

 \bibitem{Weinberg}
S. Weinberg, Phys. Rev. Lett. {\bf 40}, 223  (1978).

 \bibitem{Wilczek}
F. Wilczek, Phys. Rev. Lett. {\bf 40}, 279  (1978).

 \bibitem{Preskill}
J. Preskill and M. B. Wise and F. Wilczek, Phys. Lett. B {\bf 120}, 127  (1983).

\bibitem{Abbott}
L. F. Abbott and P. Sikivie, Phys. Lett. B {\bf 120}, 133  (1983).

\bibitem{Dine}
M. Dine and W. Fischler, Phys. Lett. B {\bf 120}, 137  (1983).

\bibitem{Davis}
R. L. Davis, Phys. Lett. B {\bf 180}, 225  (1986).

\bibitem{Brookhaven}
D. M. Lazarus and G. C. Smith and R. Cameron and A. C. Melissinos and G. Ruoso and Y. K. Semertzidis and F. A. Nezrick, Phys. Rev. Lett. {\bf 69}, 2333  (1992).

\bibitem{Japan}
S. Moriyama and M. Minowa and T. Namba and Y. Inoue and Y. Takasu, Phys. Lett. B {\bf 434}, 147  (1998).

\bibitem{SOLAX}
F. T. Avignone III {\sl et al.}, (SOLAX Collaboration), Phys. Rev. Lett. {\bf 81}, 5068  (1998).

\bibitem{DAMA90CL}
R. Bernabei {\sl et al.}, Phys. Lett. B {\bf 515}, 6  (2001).

\bibitem{Japan2}
Y. Inoue and T. Namba and S. Moriyama and M. Minowa and Y. Takasu and T. Horiuchi and A. Yamamoto, Phys. Lett. B {\bf 536}, 18  (2002).

\bibitem{COSME}
A. Morales {\sl et al.}, (COSME Collaboration), Astropart. Phys. {\bf 16}, 325  (2002).

\bibitem{ADMX}
S. J. Asztalos {\sl et al.}, (ADMX Collaboration), Phys. Rev. D  {\bf 69}, 011101  (2004).

\bibitem{CAST}
K. Zioutas {\sl et al.}, (CAST Collaboration), Phys. Rev. Lett. {\bf 94}, 121301  (2005).

\bibitem{CASTvacuum}
S. Andriamonje {\sl et al.}, (CAST Collaboration), J. Cosmol. Astropart. Phys  {\bf 04}, 010  (2007).

\bibitem{CDMS}
Z. Ahmed {\sl et al.}, (CDMS Collaboration), Phys. Rev. Lett. {\bf 103}, 141802  (2009).

\bibitem{EDELWEISS}
E. Armengaud {\sl et al.}, J. Cosmol. Astropart. Phys. {\bf 11}, 067  (2013).

\bibitem{CASTBufferGas}
M. Arik {\sl et al.}, (CAST Collaboration), Phys. Rev. Lett. {\bf 112}, 091302  (2014).

\bibitem{Creswick}
R. J. Creswick and F.T. Avignone III and H.A. Farach and J.I. Collar and A.O. Gattone and S. Nussinov and K. Zioutas, Phys. Lett. B {\bf 427}, 235  (1998).

\bibitem{CreswickPRD}
R. J. Creswick and S. Nussinov and F.T. Avignone III, Phys. Rev. D {\bf 78}, 017702  (2008).

\bibitem{Raffelt}
G. G. Raffelt, Phys. Repts. {\bf 198}, 1  (1990).

\bibitem{Sikivie}
P. Sikivie, Phys. Rev. Lett. {\bf 51}, 1415  (1983).

\bibitem{Bibber}
K. van Bibber and P. M. McIntyre and D. E. Morris and G. G. Raffelt, Phys. Rev. D {\bf 39}, 2089  (1989).

\bibitem{Hoogeveen}
W. Buchm\"{u}ller and F. Hoogeveen,  Phys. Lett. B {\bf 237}, 278  (1990).

\bibitem{Zioutas}
E. A. Paschos and K. Zioutas,  Phys. Lett. B {\bf 323}, 367  (1994).

\bibitem{CUORE}
C. Arnaboldi {\sl et al.}, (CUORE collaboration), Nucl. Instrum. Meth. A {\bf 518}, 775  (2004).

\bibitem{CUOREproposal}
R. Ardito {\sl et al.}, (CUORE collaboration), hep-ex/0501010, (2005).

\bibitem{TeO2}
C. Arnaboldi {\sl et al.}, Journal of Crystal Growth {\bf 312}, 2999  (2010).

\bibitem{DFTtheorem}
P. Hohenber and W. Kohn, Phys. Rev {\bf 136}, B864  (1964).

\bibitem{DFTequation}
W. Kohn and L. J. Sham, Phys. Rev {\bf 140}, A1133  (1965).

\bibitem{WIEN2k}
P. Blaha and K. Schwarz and G.K.H. Madsen and D. Kvasnicka and J. Luitz, WIEN2k: An Augmented Plane Wave + Local Orbitals Program for Calculating Crystal Properties (Vienna University of Technology, Vienna, Austria), (2001)


\bibitem{BGrate}
F. Alessandria {\sl et al.}, J. Cosmol. Astropart. Phys {\bf 038}, 1475 (2013).

\bibitem{KSVZ}
J. E. Kim,  Phys. Rev. Lett. {\bf 43}, 103  (1979).

\bibitem{KSVZ1}
M. A. Shifman and A. I. Vainshtein and V. I. Zakharov, Nucl. Phys. B  {\bf 166}, 493  (1980).

\bibitem{DFSZ}
A. R. Zhitnitsky,  Sov. J. Nucl. Phys  {\bf 31}, 260  (1980).

\bibitem{DFSZ1}
M. Dine and W. Fischler and M. Srednicki,  Phys. Rev. Lett. {\bf 104}, 199 (1981).

\bibitem{IAXO}
J. K. Vogel {\sl et al.}, Physics Procedia  {\bf 61}, 193  (2015).

\end{thebibliography}
\end{document}